\documentclass[letterpaper, 10 pt, journal, twoside]{ieeetran}

\IEEEoverridecommandlockouts                              

\usepackage{stfloats}
\usepackage[utf8]{inputenc}
\usepackage{graphicx} 
\usepackage{color}
\usepackage{comment}
\usepackage{amssymb} 
\usepackage{amsmath}
\usepackage{multirow}
\usepackage{float}
\usepackage{bm}
\usepackage{cancel}
\usepackage{subfigure}
\usepackage[dvipsnames]{xcolor}
\usepackage{cite}
\usepackage{booktabs}

\usepackage[top=2.01cm, bottom=1.52cm, left=1.69cm, right=1.69cm]{geometry}
\usepackage[absolute,overlay]{textpos} 

\usepackage{algorithm}
\usepackage[noend]{algpseudocode}

\title{\LARGE \bf
Thermally-Resilient Soft Gripper for On-Orbit Operations}

\author{F. Ruiz$^{1}$, B.C. Arrue$^{1}$ and A. Ollero$^{1}$
\thanks{Manuscript submitted: March 15th, 2024.}
\thanks{This work has been developed within the framework of the AERO-TRAIN project, a Marie-Sklodowska-Curie ITN. The funding of AROSA (Smart Robotics for On-Orbit Servicing Applications) is also acknowledged.}
\thanks{$^{1}$ F. Ruiz (frvincueria@us.es), B.C. Arrue
(barrue@us.es) and A. Ollero
(aollero@us.es) are with the GRVC Robotics Lab of Seville, Spain.
}%
}

\begin{document}

\maketitle

\begin{abstract}

Research in soft manipulators has significantly enhanced object grasping capabilities, thanks to their adaptability to various shapes and sizes. Applying this technology to on-orbit servicing, especially during the capture and containment stages of active space debris removal missions, might offer a secure, adaptable, and cost-effective solution compared to the trend of increasing the degrees of freedom and complexity of the manipulator (e.g. ClearSpace, Astroscale). This work aims to conduct an experimental proof of concept, for which challenges such as radiation, vacuum, and microgravity are significant, but the predominant issue is ensuring effective operation in the extreme temperature swings, where flexible materials may exhibit cryogenic crystallization or drastic shifts in their elasticity. This work addresses this challenge through an initial stage of analytical modeling of the thermal dynamics inside the manipulator in orbit; which is then used for the development of a first experimental prototype tested with liquid nitrogen and heat guns. The multi-layered design for Low Earth Orbit (LEO) leverages the properties of TPU at low infill rates for lightweight inherent flexibility, silicone rubber ensuring structural integrity, PTFE (Teflon) for unparalleled thermal stability, and aerogel for insulation. The tendon-actuated servo-driven gripper is tested in the laboratory by varying the shape and size of objects during the grasping. The results, based on servomotor force metrics to assess the flexible manipulator's adaptability and object capture efficiency across temperature changes, affirm the concept's viability. Forces increase up to 220$\%$ in cryogenic conditions and decrease by no more than 50$\%$ at high temperatures.


\end{abstract}

\vspace{-3.5mm}
\section{INTRODUCTION}

\IEEEPARstart{S}{pace} exploration has always pushed the boundaries of technology, with robots playing an increasingly vital role. Historically, space robots have been rigid in structure. While these rigid robots have proven effective in many missions, their limitations become evident in challenging terrains, as seen with the Spirit rover, which encountered difficulties in loose soil \cite{planetarywheels}. 

In contrast, soft robots, typically crafted from materials with a Young's modulus under 1 GPa, present numerous advantages. They offer more flexibility and degrees of freedom, which enhances their adaptability to diverse space environments \cite{Rus2015DesignFA}. Notably, soft robots can better manage collisions, minimizing damage and thus outperforming rigid counterparts in specific scenarios. Flexibility in space technology has already been explored in various forms, such as deployable solar arrays \cite{solararray} and inflatable space modules \cite{inflatables}, among others \cite{scienceHawkes}.

Recent proposals have introduced innovative concepts of soft robots for space exploration \cite{softspace}, including deployable Mars rovers \cite{pop_up_mars}, satellite grippers \cite{satellite_gripper}, and others. Although these concepts harness adaptability suitable for varied space missions, the extreme operational conditions in space environments can pose challenges. The development of space-ready soft robots is still in early stages. For example, the silicone-based satellite gripper in \cite{satellite_gripper} might face severe challenges in the low temperatures of a LEO orbit. Even higher TRL concepts, some trialed on the ISS such as the gecko-inspired gripper \cite{gecko}, recognize temperature adaptation as a key hurdle for prototypes.

\begin{figure}
\includegraphics[width=0.492\textwidth,scale=0.25]{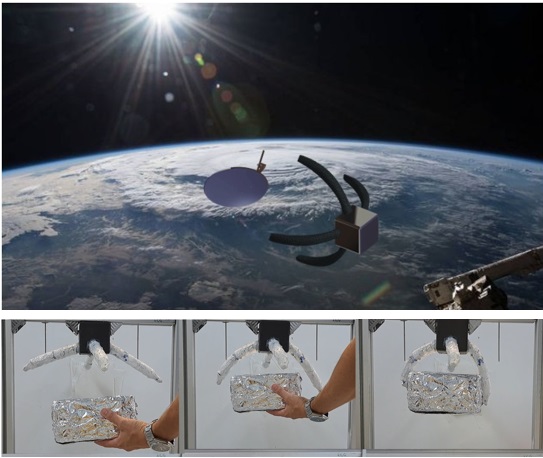}
\caption{A thermally-resilient soft robotic manipulator proof of concept designed for the capture and containment stages in active space debris removal missions in Low Earth Orbit}
\label{figure1}
\vspace{-5mm}
\end{figure}

To develop space-qualified soft robotics, it's vital to bridge the gap between robotics, materials engineering, and cryogenics. Teflon (PTFE), which can display flexibility under certain conditions, is notable for its broad temperature range (-260°C to 250°C) and has been foundational in space-related components due to its thermal properties \cite{teflon1, teflon2}. However, most materials used in soft robotics are not suitable for space because of their thermal limitations, making the study of elastomers an ongoing and active field. Silicones, as an example, can resist crystallization even at -120°C \cite{silicone1}, and their behavior at elevated temperatures is well-understood \cite{silicone2}. The performance of TPU in frigid conditions is under investigation \cite{TPU1}, with its operational range believed to be between -90°C and 130°C \cite{TPU2}. While aerogels have demonstrated excellent thermal endurance \cite{silica_aerogel_1}, they may become brittle in certain configurations. Molybdenum grease is renowned as an optimal cryogenic lubricant \cite{molibdeno1}.

Addressing the aforementioned challenge is paramount due to the impact that space-qualified soft manipulators can have on the pressing issue of space debris. Initially conceived in the 1980s for satellite servicing \cite{manipulators_space}, space manipulators have now become essential for debris removal. The first active debris removal mission is scheduled for 2025, a public-private collaboration between ESA and ClearSpace \cite{Clearspace}. Currently, we have around 20,000 large debris pieces in Low Earth Orbit (LEO) \cite{esa_debris}, in addition to numerous smaller fragments, posing a significant collision threat. In fact, large satellites in LEO confront a yearly 0.01$\%$ collision risk with sizable debris, a percentage that increases as the debris size decreases. Such impacts can multiply the debris count, exemplifying the Kessler Syndrome and threatening the future of space exploration.

Efforts to address this problem have been varied to date, not only involving active debris removal with robotic arms \cite{manipulators_space}. Other strategies include the use of large nets \cite{net_debris}, tether systems \cite{tether_debris}, laser ablation \cite{laser_debris}, and magnetic methods \cite{magnetic_debris}. Drag devices are also in play, aiming to expedite debris re-entry into Earth's atmosphere. However, as previously mentioned, recent research emphasizes the capabilities of soft robotic manipulators for active debris removal. For example, Stanford's gecko-inspired method leans on the adhesive qualities of the reptile's feet \cite{gecko, gecko2}. Other soft robotic manipulators have been proposed \cite{softdebris, softdebris2}, while the aforementioned extreme conditions are still a major challenge \cite{softspace}.

\begin{figure*}
\includegraphics[width=0.97\textwidth,scale=0.25]{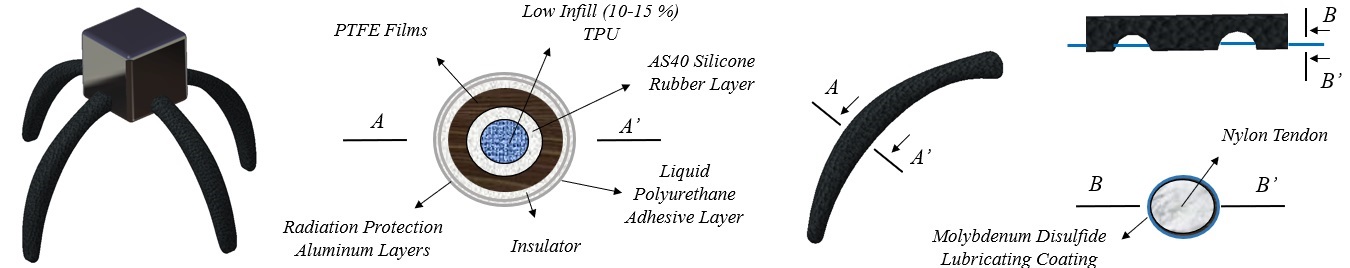}
\caption{Cross-sectional view of the multi-layered soft gripper design, detailing each specific layer's material. }
\label{figure1}
\vspace{-4mm}
\end{figure*}

This work conducts experiments using liquid nitrogen to mimic cryogenic conditions and heat guns to replicate elevated temperatures. These tests offer insights into material behavior under such conditions, guiding the development of a soft gripper designed specifically for space debris removal in LEO orbits. The gripper's multi-layered structure incorporates thermoplastic polyurethane for flexibility at low infill rates, silicone rubber for structural support, PTFE (Teflon) for exceptional thermal resistance, and aerogel for insulation. Additionally, a tendon-driven system, made of nylon and lubricated with molybdenum disulfide, ensures functionality in freezing conditions. This gripper demonstrates consistent performance across a temperature range of -180°C to 220°C.

To ensure a clear understanding of this work, the structure of this article is organized as follows: Section 2 outlines the challenges associated with operating soft robots in orbit. Section 3 covers the thermal modeling of the system. Section 4 details the proposed high-level design of the manipulator. Section 5 simulates the multi-layered design and optimizes each layer thickness. Section 6 combines a description of the experimental setup with an analysis of the results. Finally, Section 6 concludes the paper, emphasizing the potential of the proposed manipulator on space debris removal initiatives.

\vspace{-2mm}

\section{Problem Statement}

Space debris, encompassing defunct satellites, spent rocket stages, and collision fragments, poses significant risks due to its unpredictability. This section focuses on the main challenges of active debris removal using manipulators, addressing both the harsh conditions of space and the difficulty of adapting to the diverse sizes and shapes of debris.

\vspace{-5mm}

\subsection{Types and Quantities of Space Debris}

Differentiating space debris by size can provide insights into the varying challenges each size group presents. Table \ref{tab:debris_size} summarizes these categories.

\vspace{-3mm}

\begin{table}[h!]
\caption{Classification of space debris by size and shape.}
\centering
\begin{tabular}{|c|c|c|}
\hline
\textbf{Size} & \textbf{Shape/Description} & \textbf{Quantity} \\
\hline
$\geq 10cm$ & Defunct Satellites, Rocket Stages & 20,000+ \\
\hline
$1cm - 10cm$ & Smaller Fragments, Bolt, Paint Flecks & 500,000+ \\
\hline
$\leq 1cm$ & Micro-debris, tiny fragments  & 100 million+ \\
\hline
\end{tabular}
\label{tab:debris_size}
\vspace{-5mm}
\end{table}

\subsection{Space Environment and Temperature Challenges}

The unique conditions of space, characterized by vacuum and microgravity, influence heat transfer dynamics profoundly. Without an atmosphere, there's no convection; thus, objects heat up dramatically when exposed to the Sun and cool down rapidly when in its shadow. Specifically, in Low Earth Orbit (LEO), a spacecraft experiences about 45 minutes of direct sunlight followed by 45 minutes of darkness. During the sunlit phase, temperatures can soar up to 220°C. But once in the shadow, they can plunge to around -200°C. This rapid thermal cycling is due to the lack of an atmosphere to retain heat, making the heat accumulated during the sunlit phase dissipate quickly.

Designing systems, especially soft robotic manipulators for space, requires thorough understanding of the aforementioned extreme conditions. Materials must retain their properties and functions across wide temperature ranges.

\section{Thermal Dynamics Modeling}

This section covers heat transfer dynamic modeling within the soft manipulator. This model will be used for simulations that will allow optimizing the layered design, taking into account material constraints. 

\vspace{-3mm}

\subsection{Heat Transfer Modeling in Orbit}

In the space environment, conduction is the primary mechanism for heat movement within the manipulator's layers due to molecular agitation. Convection is generally non-existent in the vacuum of space, radiative heat transfer takes precedence. The proposed model considers both conductive and radiative heat transfer in a transient manner. The governing equation for time-dependent conduction in cylindrical coordinates is:

\begin{equation}
\rho c_p \frac{\partial T}{\partial t} = \frac{1}{r} \frac{d}{dr} \left( k \frac{dT}{dr} \right) + q_{\text{radiation}}
\end{equation}

Where:
\begin{itemize}
    \item \( \rho \) is the material density.
    \item \( c_p \) is the specific heat capacity at constant pressure.
    \item \( T \) is the temperature.
    \item \( r \) is the radial coordinate.
    \item \( k \) is the material thermal conductivity.
    \item \( q_{\text{radiation}} \) represents the heat added or removed due to radiative transfer.
\end{itemize}

\subsection{Numerical Solution and Discretization}

To solve the equation, a finite difference method (FDM) is employed. For the radial derivative, a central difference scheme can be applied:
\begin{equation}
\frac{dT}{dr} \approx \frac{T_{i+1} - T_{i-1}}{2 \Delta r}
\end{equation}

The second radial derivative can be discretized similarly using the central difference scheme, leading to:
\begin{equation}
\frac{d^2T}{dr^2} \approx \frac{T_{i+1} - 2T_i + T_{i-1}}{\Delta r^2}
\end{equation}

The temporal discretization can be achieved using a forward difference for the time derivative:
\begin{equation}
\frac{\partial T}{\partial t} \approx \frac{T_i^{n+1} - T_i^n}{\Delta t}
\end{equation}

Where \( n \) is the time step index.

The stability of the numerical solution is crucial. In this context, the CFL condition dictates the relationship between the time step \( \Delta t \) and the spatial step \( \Delta r \) to ensure a stable and accurate solution:

\begin{equation}
\alpha \frac{\Delta t}{(\Delta r)^2} \leq \frac{1}{2}
\end{equation}

where $\alpha$ is the thermal diffusivity.

\subsection{LEO Orbit Properties}

The soft manipulator is designed for operations in a Low Earth Orbit (LEO). The properties of this orbit, typical for LEO altitudes, are summarized in Table \ref{tab:LEO_properties}. It's noteworthy that the radiation value provided is an approximation of the radiation received in the LEO orbit.

\begin{table}[h]
\centering
\caption{Properties of the LEO Orbit}
\label{tab:LEO_properties}
\begin{tabular}{|c|c|}

\hline
\textbf{Parameter} & \textbf{Value} \\
\hline
Orbit Altitude & 160 km - 2,000 km \\
Orbit Duration & 90 minutes \\
Time in Sunlight & 45 minutes \\
Time in Shadow & 45 minutes \\
Radiation Value (Approx.) & \( q_{\text{rad}} = 1460 \) W/m\(^2\) \\
\hline
\end{tabular}

\vspace{-5mm}
\end{table}

\section{High-Level Soft Manipulator Design}

This section introduces the high-level design of the soft gripper and the proposed materials, which will be simulated and optimized in Section V. Figure 2 offers an in-depth illustration of the gripper's design, including a cross-sectional view that highlights the various materials used.

\vspace{-2mm}

\subsection{Material Selection and Multi-Layered Approach}

The designed gripper utilizes a multi-layered structure:

\begin{itemize}
\item \textbf{Inner Layer (TPU):} Thermoplastic polyurethane (TPU) is chosen due to its lightweight and flexibility, especially at low infill rates of 10-15$\%$. However, its use throughout the entire manipulator is restricted since it weakens considerably at high temperatures and becomes brittle at extreme lows. Positioned at the core, it remains shielded from the most drastic temperature swings in space.

\item \textbf{Intermediate Structural Layer (Silicone):} Silicone, as an elastomer, adds critical structural integrity. However, a combined approach with TPU is adopted instead of using silicone exclusively. This is not only to benefit from the lightweight nature of TPU at low infill rates, thereby reducing the overall weight, but also to minimize inherent vibrations observed in silicone. Such vibrations could pose operational challenges in space. Furthermore, the experiments have shown that silicone crystallizes below -120°C, an even lower limit than TPU, so it is consistent to place it in a more outer layer.

\item \textbf{External Thermal Stability Layer (PTFE):} PTFE, known for its minimal elasticity variations (around 10$\%$) across a wide temperature range (-200°C to 220°C), is selected for the external layer. While it's not ideal for the entirety of the gripper due to its higher modulus of elasticity at room temperature, its thermal stability makes it apt as an external layer to provide thermal resilience, especially at low infill rates of 25-30\%. However, its manufacturing challenges will be discussed further.

\item \textbf{Thermal Insulation Layer:} A thin layer of flexible aerogel is introduced for its superior insulation properties, acting as a buffer to shield the manipulator's internals from extreme temperatures.

\item \textbf{Protective and adhesion films:} The outermost layer is made of reflective aluminum, which can reduce radiation effects. Finally, liquid polyurethane enhances the gripper's bond with the object, ensuring a firm grip and minimizing the risk of slip during operations.

\end{itemize}

This multi-layered approach strategically combines the advantages of each material to produce a gripper that's both functional and durable in the challenging space environment.

\subsection{Manufacturing Process of the Soft Gripper and Actuation}

The fabrication of the soft gripper is a multi-step procedure. Each step is detailed below and further detailed in the Supplementary Material:

\begin{enumerate}
    \item \textbf{TPU Layer Creation:} The initial layer, made from Thermoplastic Polyurethane (TPU), is produced using Fused Deposition Modeling (FDM) 3D printing at low infill rates. 
    
    \item \textbf{Silicone Layer Formation:} A mold, which can be 3D printed, is employed to shape the subsequent silicone layer. AS40 silicone rubber is mixed with a catalyst to initiate curing. Once poured into the mold, it's left to cure for a duration of 4-6 hours.
    
    \item \textbf{PTFE Layer Attachment:} Pure PTFE fabrication remains a challenge, primarily due to its high melting point. While some experimental endeavors have made headway in 3D printing PTFE, these methods are still not robust. Other techniques, such as cold molding and sintering, are complex and less feasible for the present purpose. To leverage the beneficial thermal properties of PTFE (Teflon) in this study, thin sheets consisting of 70\% PTFE and 30\% PETG are used. These sheets are layered onto the soft gripper to form the required thermal resilient shell.
    
    \item \textbf{Aerogel Insulation:} The aerogel is applied as thin membranes, ensuring uniform coverage and offering thermal insulation properties.
    
    \item \textbf{Aluminum Film and Coating:} A thin reflective aluminum paper film is adhered to the surface of the gripper to offer protection against radiation. The adhesive layer, ensuring grip with space debris, is applied using a brush.
    
\end{enumerate}

The actuation mechanism is shown in Figure 2. It is tendon-actuated, servo-driven using nylon from fishing lines. Nylon tendons maintain robustness over a wide range of temperatures. For smooth operation in cold conditions, a coating of molybdenum disulfide grease is applied using a brush for lubrication.

\begin{figure}
\includegraphics[width=0.492\textwidth,scale=0.25]{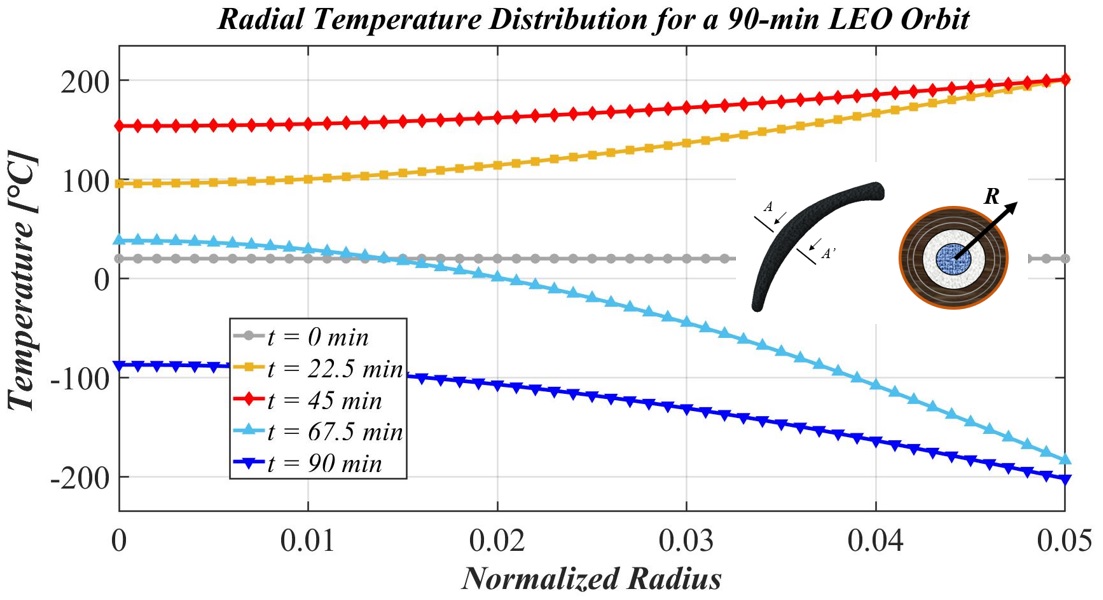}
\caption{Simulated temporal evolution of the temperature along the radius of the final design of the soft gripper throughout a complete LEO orbit cycle. The curves represent different time instances, showcasing thermal fluctuations as the manipulator transitions through sunlit and shadowed regions of the orbit.}
\label{figure1}
\vspace{-4mm}
\end{figure}

\vspace{-1mm}

\begin{figure}
\includegraphics[width=0.492\textwidth,scale=0.25]{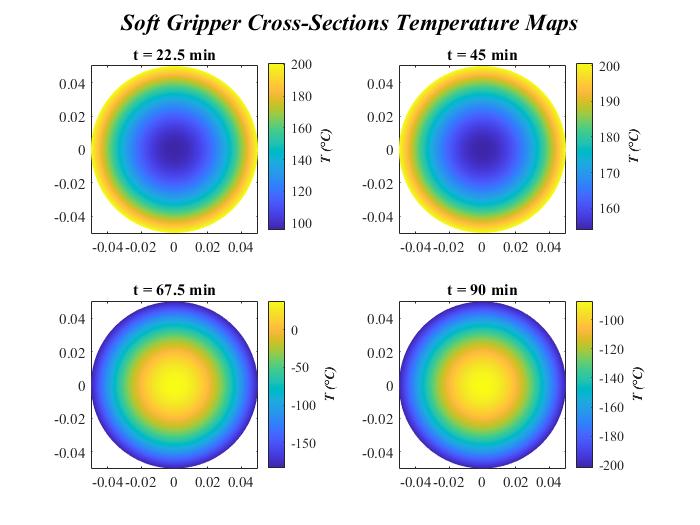}
\caption{Simulated cross-sectional temperature maps for the final design of the soft gripper at distinct moments during its LEO orbit cycle. The visualizations emphasize the temperature gradients within the structure. }
\label{figure1}
\vspace{-4mm}
\end{figure}

\section{Model Results and Final Design}

The design criterion is that each material layer remains at a safety factor of 1.2 away from its critical crystallization temperature and its melting point. This leads to the analytical result shown in Table III. For these simulations, a simplification of constant diameter in the gripper corresponding to 10cm, consistent with the following experimental prototype, was used. The iterative process produced temperature profiles (Figures 3 and 4). It is observed that the thickest layer is made of Teflon, which will reduce the manipulator's flexibility in exchange for maintaining temperature resistance.

\begin{table}[h]
\centering
\caption{Thermal properties and design choices of materials used in the soft manipulator}
\begin{tabular}{|c|c|c|c|c|}
\hline
\textbf{Material} & \textbf{Heat Coeff \( k \)} & \textbf{Density \( \rho \)} & \textbf{Specific Heat \( c_p \)} & \textbf{Layer} \\
\hline
TPU & 0.22 W/m·K & 120 kg/m³ & 1.2 J/g·K & 26 $\%$ \\
Silicone & 0.2 W/m·K & 970 kg/m³ & 1.5 J/g·K & 23 $\%$ \\
PTFE & 0.18 W/m·K & 660 kg/m³ & 1.0 J/g·K & 32 $\%$ \\
Aerogel & 0.02 W/m·K & 200 kg/m³ & 1.0 J/g·K & 19 $\%$ \\
\hline
\end{tabular}
\vspace{-3mm}
\end{table}

Moreover, thermal modeling offers key insights into the behavior of the soft manipulator under space conditions:

\begin{enumerate}
\item The radial temperature profile (Figure 3) exhibits pronounced variations. The external surface encounters temperatures fluctuating from 205°C during sunlight exposure to a chilling -197°C when in the shadow.
\item Internally, the gripper experiences a milder temperature swing, with values oscillating between 153°C and -96°C.
\item The observations highlight the crucial significance of the thermal design, proving its functionality across diverse thermal conditions.
\end{enumerate}

\begin{figure}[t]
\includegraphics[width=0.46\textwidth,scale=0.25]{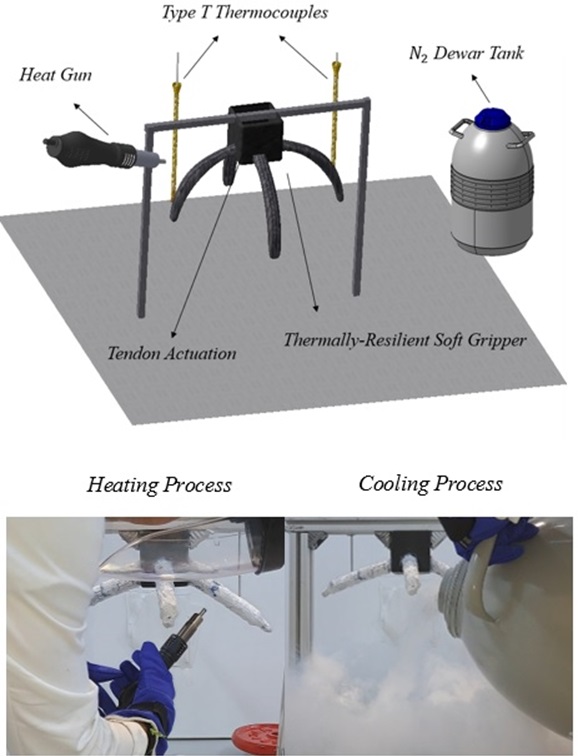}
\caption{Test bench setup: manipulator, sensing, and temperature actuators (liquid nitrogen and heat guns).}

\label{figure1}
\vspace{-4mm}
\end{figure}

\section{Experimental Prototype}

This section describes the experiments conducted to test the performance, flexibility, and response of the soft manipulator under different temperature conditions.

\vspace{-2mm}

\subsection{Test Bench and Instrumentation}

Figure 5 illustrates the test bench setup, which includes:

\begin{itemize}
\item A robust bar structure to anchor the soft manipulator.
\item A Dewar tank filled with liquid nitrogen maintained at -196°C. Immersion of the gripper into this liquid nitrogen can lower temperatures to approximately -180°C.
\item A heat gun capable of reaching temperatures up to 220°C or even higher.
\item A Type-T thermocouple, chosen for its accuracy in monitoring different ranges of temperatures.
\item The high-torque Hitec D950TW servos, utilized to regulate the force applied by the manipulator's tendons.
\end{itemize}

\vspace{-3mm}

\subsection{Manipulator Configurations and Adaptability}

The versatility of the soft manipulator was tested through its ability to handle different debris types (Figure 6). Note that the objects used for the tests have reduced mass, since in microgravity conditions the weight force is negligible (although not the inertia).

\begin{itemize}
    \item \textbf{Dual Arm Configuration:} Using just two of its arms, this setup is suitable for simpler debris (such as complete satellite elements with regular shapes, or small rocket stages), ensuring a firm grasp.
    
    \item \textbf{Quad Arm Configuration:} By using all four arms, this setup is designed for irregularly shaped debris (as loose fragments or pieces in orbit after collisions), in which the previous configuration may not be sufficient and cause the object to slide laterally, something essential to avoid in space in microgravity conditions due to the energy expenditure that would be required to recover its position. This configuration provides greater stability and control. 
\end{itemize}

\begin{figure}[t]
\includegraphics[width=0.492\textwidth,scale=0.25]{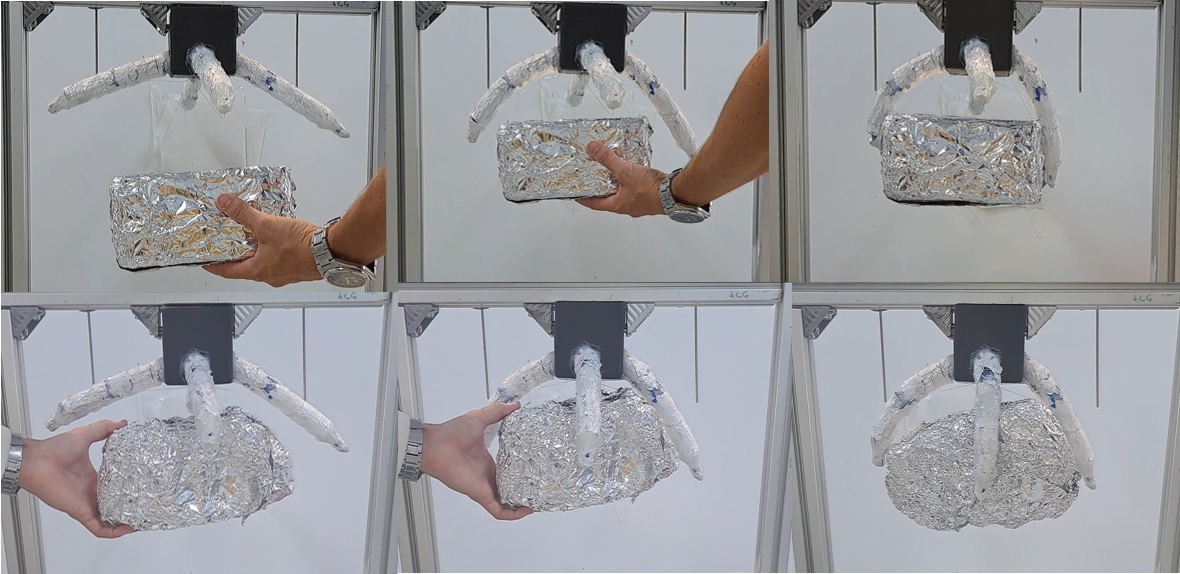}
\caption{Demonstration of the soft gripper's versatile gripping capabilities. Top: Dual-arm configuration showcasing the adaptability in handling objects. Bottom: Four-arm approach emphasizing the gripper's enhanced stability and grasp strength when dealing with larger or more cumbersome objects.}

\label{figure1}
\vspace{-4mm}
\end{figure}

In Table IV, the evolution of the force required by the servomotor based on the shape of the object and the gripper configuration (either dual or quad arm) is observed. Although the total force required by the servomotor is always greater in the quad arm case, the force per arm significantly decreases for irregular objects (in these types of objects, the contact area is smaller). This allows for achieving a stable grip without exerting as much pressure. In addition to making operations in space safer due to previously discussed microgravity reasons, this configuration may even be more energy-efficient.

\vspace{-3mm}

\begin{table}[h]
\centering
\caption{Force requirements for stable gripping of different debris and arm configurations.}
\label{tab:forceRequirements}
\begin{tabular}{|c|c|c|c|}
\hline
\textbf{Debris Shape} & \textbf{Metrics} & \textbf{Dual Arm} & \textbf{Quad Arm} \\
\hline
\multirow{2}{*}{\textbf{Regular}} & Total Force (kgcm) & 7.8 & 13.2 \\
\cline{2-4}
& Force per Arm (kgcm) & 3.9 & 3.3 \\
\hline
\multirow{2}{*}{\textbf{Irregular}} & Total Force (kgcm) & 12.3 & 14.4 \\
\cline{2-4}
& Force per Arm (kgcm) & 6.1 & 3.6 \\
\hline
\end{tabular}
\vspace{-3mm}
\end{table}

\vspace{-3mm}

\subsection{Temperature Variation Performance Metrics and Results}

The thermal experiments offer insights into the servo's force requirements over a range of temperatures. As depicted in Figure 7, at low temperatures, the required force doubled due to material stiffening. If the soft gripper were made solely of silicone, this increase would be much greater due to crystallization (3 or 4 orders of magnitude higher in the elasticity modulus). On the other hand, as temperatures approached 200°C, the necessary force halved. This effect could be much more notable if closer to the melting point, but the design criterion allows for controlled deformations to be maintained.

Repeated thermal cycling accentuated these variations (see Figure 8). The force rose by as much as 35$\%$ under cold conditions, but diminished by 18$\%$ in the hot configuration. The force stayed largely consistent at ambient temperatures.

\vspace{-4mm}

\section{Conclusions}

This paper presented the experimental proof of concept of a soft gripper tailored for space debris removal in Low Earth Orbit (LEO). The multi-layered, simulated and validated design strategically integrates elastomeric materials such as TPU and silicone, which provide flexibility, combined with the thermal stability of PTFE and aerogel insulation, leading to a thermally-resilient gripper. 

Notably, the gripper's external surface can experience temperature variations from 205°C to -197°C in LEO orbits, while the design maintains safer internal temperature boundaries. Experimental evaluations highlighted the gripper's ability to grasp objects of diverse shapes, responding effectively to the multifaceted nature of space debris. Observations from the test bench emphasized the influence of harsh temperatures on the servo's force requirements: at cryogenic temperatures, hardening was observed, while at higher temperatures, it became more flexible; both effects becoming more pronounced with the number of cycles.

\begin{figure}
\includegraphics[width=0.492\textwidth,scale=0.25]{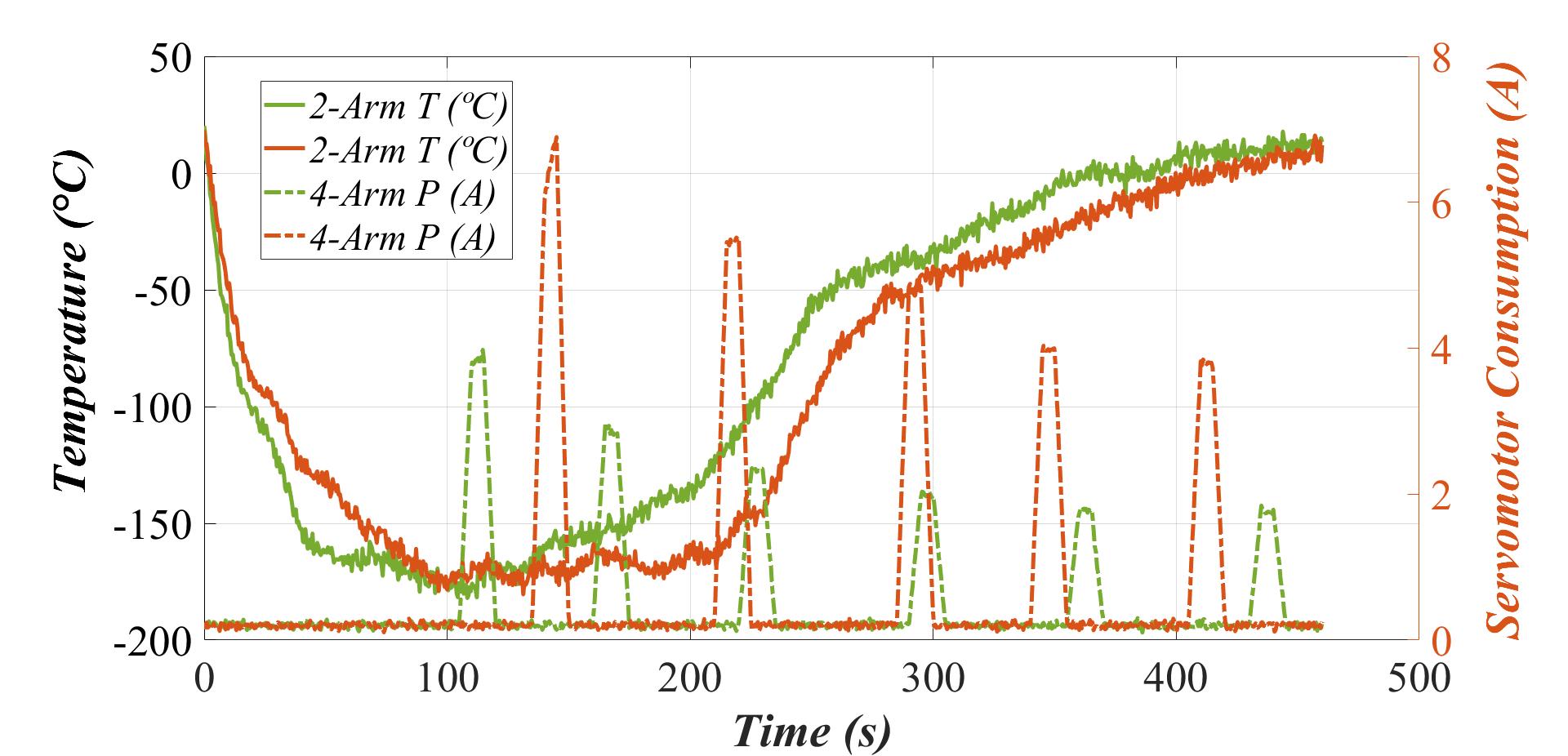}
\caption{Experimental results for a temperature cycle (dual arm and quad arm, respectively, under regular debris configuration) showing servomotor consumption in different gripping cases }
\label{figure1}
\vspace{-3mm}
\end{figure}

This work has aimed to bridge the gap between soft robotics and space engineering, through pioneering experimentation in this field with liquid nitrogen. Future work will include more extensive experimentation, involving the use of thermal cameras for temperature measurement, tests under vacuum conditions, optimization of the manufacturing techniques for the Teflon layer, and improvement of the grasping technique.

\begin{figure}
\includegraphics[width=0.492\textwidth,scale=0.25]{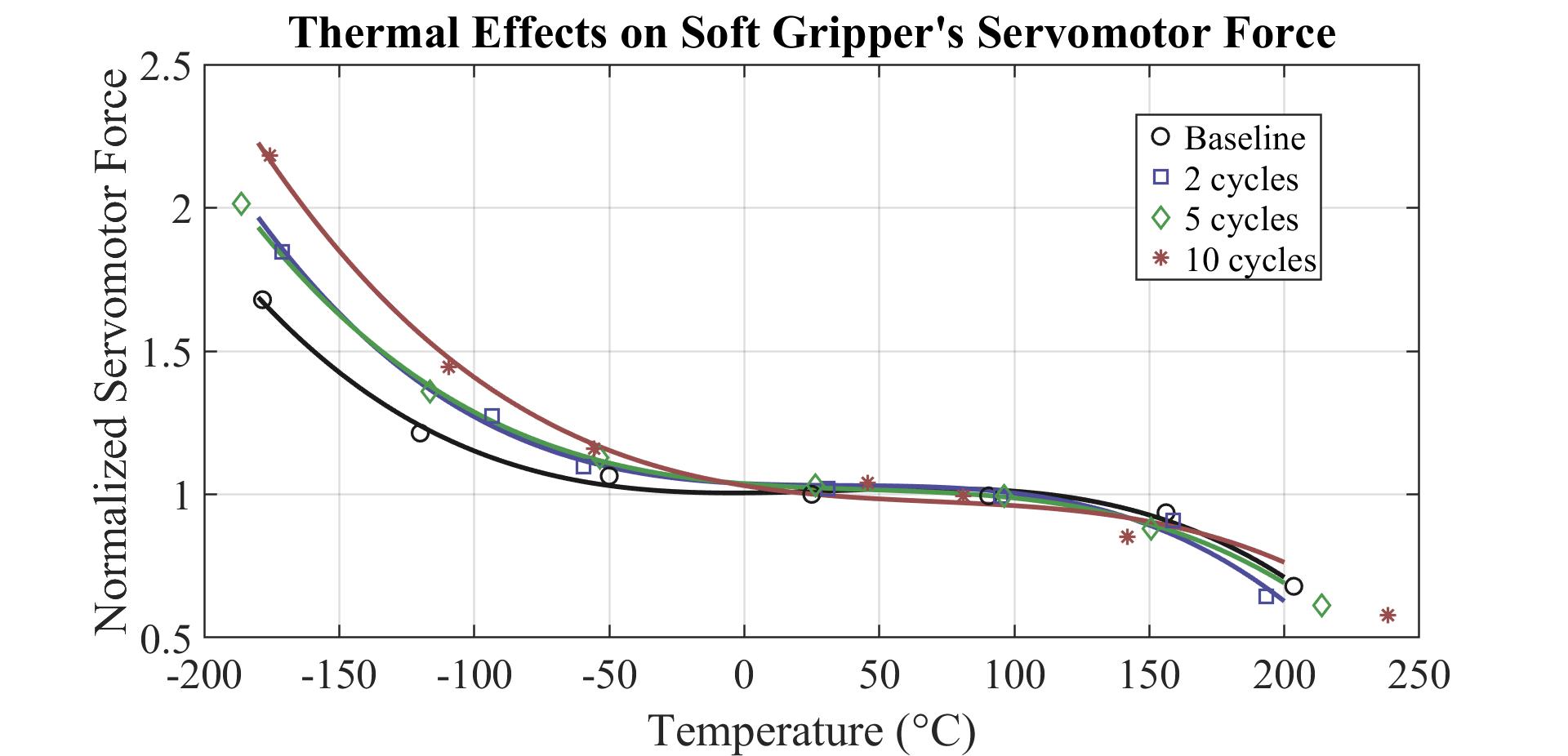}
\caption{Servomotor force variation (dual arm, regular debris shape) of the soft gripper prototype with temperature. The baseline indicates initial conditions, with subsequent symbols showing the effects after 2, 5, and 10 thermal cycles. }
\label{figure1}
\vspace{-5mm}
\end{figure}



\vspace{-3mm}

\section*{ACKNOWLEDGMENTS}
We thank Robotics, Vision and Control Group (GRVC).


\vspace{-3mm}

\bibliographystyle{IEEEtran}
\bibliography{References}

\end{document}